\documentclass[twoside]{dis08}
\usepackage[latin1]{inputenc}
\usepackage[dvips]{graphicx,epsfig,color}
\usepackage{wrapfig,rotating}
\usepackage{amssymb,amsmath,array}

\newcommand{\dels}{\delta q^s}
\newcommand{\delv}{\delta q^v}

\newcommand{\bi}{\scriptstyle b_{1}}
\newcommand{\biNN}{\scriptscriptstyle b_{1}NN}
\newcommand{\hi}{\scriptscriptstyle h_{1}}
\newcommand{\hiNN}{\scriptscriptstyle h_{1}NN}

\newcommand{\be}{\begin{eqnarray}}
\newcommand{\ee}{\end{eqnarray}}

\begin{document}
\title{Exclusive Pion-production and the Relation to Transversity}

\author{Gary R. Goldstein$^1$ and Simonetta Liuti$^2$
\thanks{This work is supported by the U.S. Department of Energy grants no. DE-FG02-01ER4120 (S.L), and no. DE-FG02-92ER40702 (G.R.G.).}
\vspace{.3cm}\\
1. Tufts University - Department of Physics \\
Medford, MA 02155 USA.
\vspace{.1cm}\\
2. University of Virginia - Department of Physics \\
Charlottesville, VA 22901 USA. \\
}

\maketitle

\begin{abstract}
Exclusive $\pi^0$ electroproduction from nucleons is suggested for extracting the tensor charge and other quantities related to transversity from experimental data. This process isolates C-parity odd and chiral odd combinations of t-channel exchange quantum numbers. In a hadronic picture it connects the meson production amplitudes to C-odd Regge exchanges with final state interactions. In a description based on partonic degrees of freedom, the helicity structure for this C-odd process relates to the quark helicity flip, or chiral odd generalized parton distributions.
\end{abstract}

We begin with a set of questions that will be answered hereafter. What is Transversity? What does transversity tell us about partons? What is the tensor charge? How can the tensor charge be measured? What is the r${\hat o}$le of exclusive scattering in accessing transversity?

Transversity itself is a concept introduced in the 1970's by Goldstein and Moravcsik~\cite{g_mor} as the appropriate spin quantization for two body scattering amplitudes, e.g. $f_{a,b;c,d}(s,t)$ that is most easily related to single spin asymmetry or azimuthal asymmetry observables. The transversity of a massive particle in its rest frame is its spin projection along an axis perpendicular to the two body scattering plane ($S\cdot p_{in} \times p_{out}$). In terms of helicities, transversity for spin  $\frac{1}{2}$ particles is $|\pm \frac{1}{2} )_T=\{|+\frac{1}{2}\rangle\pm (i)|-\frac{1}{2}\rangle\}/\sqrt{2}$.

The tensor charge is the first moment or the norm of the parton transversity distribution, $h_1(x)$. It is defined as the transversely polarized nucleon matrix element of local quark field operators,
$$
\langle P,S_T|\bar{\psi}\sigma^{\mu\nu}\gamma_5 \frac{\lambda^a}{2} \psi |P,S_T\rangle
=2\delta q^a (\mu^2) (P^\mu S_T^\nu- P^\nu S_T^\mu).$$
Like other charges, it is the integral of a distribution $(\delta q^a(x)-\delta \bar{q}^a(x))$, where  $\delta q^a(x)=h_1^a(x)$ is the {\bf transversity distribution}. It is essentially the probability to find a transversity $+\frac{1}{2}$ quark in a nucleon of transversity $+\frac{1}{2}$. Unlike the longitudinal distribution $g_1(x)$, $h_1(x)$ receives no contributions from gluons.

A question in hadronic physics is how the tensor charge can be determined, theoretically and experimentally. Some predictions and fits from various processes give:

QCD Sum rules (He and Ji,~\cite{heji})    $\delta u= 1.26, \delta d = -0.17$

Lattice (QCDSF, M. Gockeler {\it et al.},~\cite{gock}) $\delta (u-d)= 1.09 ± 0.02$

Phenomenological (Anselmino {\it et al.},~\cite{anselm}) $\delta u= 0.48\pm0.09, \delta d=-0.62\pm0.30$

Axial vector dominance(Gamberg and Goldstein,~\cite{g_g}) $\delta u= 0.58\pm0.20, \delta d =-0.11\pm0.20.$

The Gamberg and Goldstein theoretical model is based on axial vector dominance by the $b_1$(1235) and $h_1$(1170) $h_1\prime$(1380), with $J^{PC}=1^{+-}$, that couple to the tensor Dirac matrix $\sigma^{\mu\nu}\gamma_5$. The Dirac matrix has C-parity minus, which will be a crucial fact. The resulting formulae for the isovector and isoscalar tensor charges are 
$$
\delv =\frac{f_{\bi}g_{\biNN}\langle k_{\perp}^2\rangle }{\sqrt{2} M_N
M_{\bi}^2}\, ,\quad
\dels = \frac{ f_{\hi}g_{\hiNN}\langle k_{\perp}^2\rangle }{\sqrt{2} M_N
M_{\hi}^2},
$$
Because the axial vector couplings involve an additional angular momentum, to obtain the tensor structure a transverse momentum enters the coupling - the pure pole term decouples at zero momentum transfer. The interpretation that was adopted was that the coupling involves the quark constituents and thus does not vanish at zero momentum transfer. The average transverse momentum thus gives non-zero results. This led to three questions. How could the quarks be explicitly represented in the interaction with the axial vectors? The approach to answering this lay in the GPDs, particularly in the ERBL region. Hence exclusive processes should be considered, where the GPDs provide a description of hard scattering from the constituents. Secondly, how could the pole at the axial mass extrapolate to the $t=0$ limit, where the tensor charge is evaluated? This suggests the  Regge pole approach, which naturally allows extrapolation from physical poles to the physical scattering region, including $t=t_{min}$, which approaches $t=0$ for asymptotic energies. Hence there is an interplay between a partonic description and a hadronic description of the tensor charge and transversity. What kind of reaction would single out the quantum numbers of the axial exchanges? For this, the exclusive photoproduction and electroproduction of $\pi^0$ or $\eta$ mesons from nucleons have C-parity odd in the t-channel and hence can accommodate the appropriate axial vector exchanges.


\begin{wrapfigure}{r}{0.5\columnwidth}
\centerline{\includegraphics[width=0.45\columnwidth]{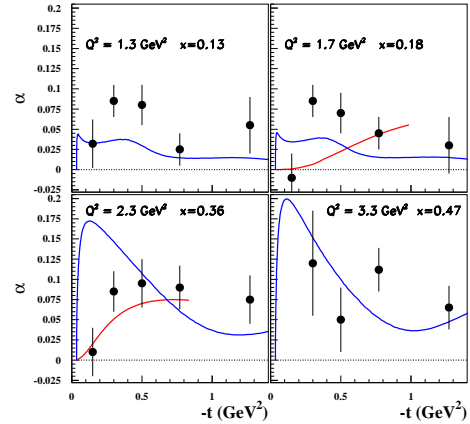}}
\caption{Beam asymmetry in Regge and GPD pictures. Data from ref~\cite{demasi}.}\label{Fig:MV}
\end{wrapfigure}

Henceforth consider these exclusive reactions, $e + p \rightarrow e^\prime + \pi^0 + p^\prime$ and the related $\eta$ and neutron target processes. The relevant subprocess is $\gamma^*+p\rightarrow \pi^0 + p^\prime$. The t-channel exchange picture involves C-parity odd, chiral odd states that include the $1^{+-}$ $b_1$ and $h_1$ mesons ($q+{\bar q}$; $S=0, L=1$ mesons) and the vector mesons, the $1^{--}$ $\rho^0$ and $\omega$ (($q+{\bar q}$; $S=1, L=0$ mesons). These axial vector mesons couple to the nucleon via the Dirac tensor $\sigma^{\mu\nu}\gamma_5$, while the vector mesons couple via $\gamma^\mu$ and/or $\sigma^{\mu\nu}$. Because of the C-parity there is no $\gamma^\mu \gamma^5$ coupling. This is quite significant in the GPD perspective - only chiral odd GPDs are involved, contrary to the accepted formulation~\cite{mank}. While ref.~\cite{mank} indicates that C-parity odd exchanges of 3 gluons, like the ``odderon", are allowed, the authors relate the process to chiral even GPDs that can involve $1^{++}$ exchange quantum numbers. This can be the case for charged pseudoscalar production, where there is not a C-parity eigenstate in the t-channel, but not for the neutral case, which has {\it definite odd C-parity}.

\begin{wrapfigure}{r}{0.5\columnwidth}
\centerline{\includegraphics[width=0.45\columnwidth]{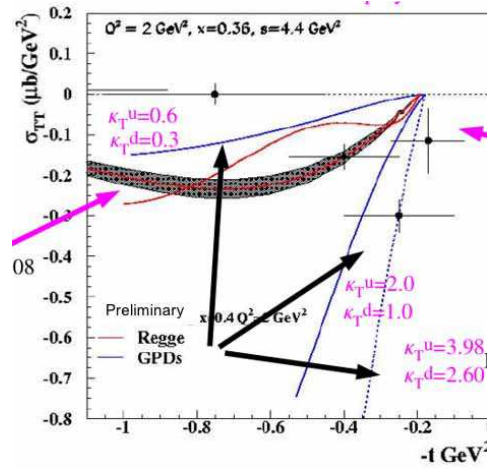}}
\caption{$d\sigma_{TT}/dt$ Regge (wavy line) and GPD pictures (with 3 sets of $\kappa_T$ pairs). Data preliminary.}\label{Fig:TT}
\end{wrapfigure}

We proceed with the hadronic picture, the Regge model for $\pi^0$ electroproduction. A successful Regge cut model was developed to fit photoproduction data many years ago~\cite{g_o}. That model essentially involves as input the vector and axial vector meson trajectories that factorize into couplings to the on-shell $\gamma+\pi^0$ vertex and the nucleon vertex. The cuts or absorptive corrections destroy that factorization, but fill in the small $t$ and $t \approx -0.5$ amplitude zeroes. To connect to electroproduction, the upper vertex factor must acquire $Q^2$ dependence. This is accomplished by replacing the elementary, t-dependent couplings with $Q^2$ dependent transition form factors. In this Regge picture the factorization for the longitudinal virtual photon is not different from the transverse photon, except for the additional power of $Q^2$ for the longitudinal case. This is in contrast to the proofs of factorization for the longitudinal case in the GPD picture, while reaction initiation by transverse photons is not expected to factorize into a handbag picture. With our form factor approach to the upper vertex (including Sudakov factors to soften the endpoint singularities) we anticipate  a similar factorization for the transverse case. 

The Regge picture is implemented by singling out the 6 independent helicity amplitudes and noting that at large $s$ and small $|t|$ the leading natural parity and unnatural parity Regge poles contribute to opposite sums and differences of pairs of helicity amplitudes. 

Now the crucial connection to the 8 GPDs that enter the partonic  description of electroproduction is through the helicity decomposition~\cite{diehl}, where, for example, one of the chiral even helicity amplitudes is given by 
$$A_{++,++}(X,\xi,t)=\frac{\sqrt{1-\xi^2}}{2}(H^q+{\tilde H}^q-\frac{\xi^2}{1-\xi^2}(E^q+{\tilde E}^q)),$$
while one of the chiral odd amplitudes is given by
$$A_{++,--}(X,\xi,t)=\sqrt{1-\xi^2}(H_T^q+\frac{t_0-t}{4M^2}{\tilde H}_T^q-\frac{\xi}{1-\xi^2}(\xi E_T^q+{\tilde E}_T^q)).$$
There are relations to PDFs, $H^q(X,0,0)=f_1^q(X)$, ${\tilde H}^q(X,0,0)=g_1^q(X)$, $H_T^q(X,0,0)=h_1^q(X)$. The first moments of these are the charge, the axial charge and the tensor charge, for each flavor $q$, respectively. Further, the first moments of $E(X,0,0)$ and $2{\tilde H}_T^q(X,0,0)+E_T^q(X,0,0)$ are the anomalous moments $\kappa^q, \kappa_T^q$, with the latter defined by Burkardt~\cite{burk}.

Chiral even GPDs have been modeled in a thorough analysis~\cite{ahlt} , based on diquark spectators and Regge behavior at small $X$, and consistent with constraints from PDFs, form factors and lattice calculations. That analysis (see the contribution of S.~Liuti in this conference) is used to obtain chiral odd GPDs via a multiplicative factor that fits the phenomenological $h_1(x)$~\cite{anselm}. With that {\it ansatz} the observables can be determined in parallel with the Regge predictions.

The differential cross section for pion electroproduction off an unpolarized target is
$$\frac{d^4\sigma}{d\Omega dx d\phi dt} = \Gamma \left\{ \frac{d\sigma_T}{dt} + \epsilon_L \frac{d\sigma_L}{dt} + \epsilon \cos 2\phi \frac{d\sigma_{TT}}{dt} 
+ \sqrt{2\epsilon_L(\epsilon+1)} \cos \phi \frac{d\sigma_{LT}}{dt} \right\}.$$
Each observable involves bilinear products of helicity amplitudes, or GPDs. For example, the cross section for the virtual photon linearly polarized out of the scattering plane minus that for the scattering plane is 
$$
\frac{d\sigma_{TT}}{dt} = \mathcal{N} \, \frac{1}{s \mid P_{CM} \mid^2} %
  2 \Re e \left( f_{1,+;0,+}^*f_{1,-;0,-} - f_{1,+;0,-}^* f_{1,-;0,+} \right).
$$

With the Regge  parameterization for the helicity amplitude that connects to $H_T(X,\xi,t)$ as $t\rightarrow t_{min}$, the tensor charge can be extracted by factoring out the meson-photon vertex (using $\Gamma(b_1\rightarrow\pi\gamma)\simeq230$ keV) and the s-dependence from the final state interaction. We obtain a preliminary result of $\delta u = 0.39\pm0.12$ and $\delta d = -0.14\pm0.04$, within the range of the global fit~\cite{anselm}. The Regge approach and the GPDs yield all observables. We show one such, the beam asymmetry (requiring $\gamma*_T-\gamma*_L$ interference) in Fig.~\ref{Fig:MV}, which is a prediction of the Regge model, given only the fit to the $Q^2=0$ data. We also show the GPD contributions, the short rising curves, for the two intermediate values of $Q^2$. All are remarkably in the range of the data. For the other observables there is preliminary data from JLab for comparison, but normalizations are still not fixed. To determine the normalization of the $Q^2$ dependence we could use available low energy parameterizations just above the resonance region. But these vary considerably. So at this time we leave the normalizations unspecified, yet expect the shapes to be reasonable. An example of $d\sigma_{TT}/dt$ is shown in Fig.~\ref{Fig:TT} where the variation in the GPD prediction is very sensitive to the values of the anomalous transverse moments. Other observables are similarly sensitive to these physical constants and, when measured, will help determine these constants~\cite{agl}.

In conclusion, we have proposed a new method, that is currently experimentally accessible, for determining fundamental constants and distributions related to the nucleon transversity, which is quite sensitive to these important quantities.

\begin{footnotesize}

\end{footnotesize}

\end{document}